\documentstyle[jgrga]{agu2001}

\authorrunninghead{HARRIS ET AL.}
\titlerunninghead{SPATIAL AND TEMPORAL VARIABILITY OF EARTH ALBEDO
$\gamma$-RAYS}

\authoraddr{Michael J. Harris, Universities Space Research Association,
1101 17th Street N.W., Suite 1004, Washington DC 20036, USA.
(harris@osse.nrl.navy.mil)}

\authoraddr{Gerald H. Share, Code 7652, Naval Research Laboratory,
Washington DC 20375-5320, USA. (share@ssd5.nrl.navy.mil)}

\authoraddr{Mark D. Leising, Department of Physics and Astronomy,
Clemson University, Box 1911, Clemson, SC 29634-1911, USA. (lmark@clemson.edu)}

\begin{document}

\title{SPATIAL AND TEMPORAL VARIABILITY OF THE GAMMA RADIATION FROM EARTH'S
ATMOSPHERE DURING A SOLAR CYCLE}

\author{Michael J. Harris} 
\affil{Universities Space Research Association, Washington DC 20036, USA}

\author{Gerald H. Share}
\affil{Naval Research Laboratory, Washington DC 20375-5320, USA}

\author{Mark D. Leising}
\affil{Department of Physics and Astronomy, Clemson University, Clemson, 
South Carolina, USA}

\begin{abstract}

The {\it Solar Maximum Mission} satellite's Gamma Ray Spectrometer 
spent much of its 1980--1989 mission pointed
at Earth, accumulating spectra of atmospheric albedo $\gamma$-rays.  Its
$28^{\circ}$ orbit ensured that a range of geomagnetic latitudes was sampled.  
We measured the variation with time and cutoff rigidity of
some key $\gamma$-ray lines which are diagnostic of the intensity
of the Galactic cosmic radiation penetrating the geomagnetic cutoff and
of the secondary neutrons produced in the atmosphere.  We found that
the intensities of nuclear lines at 1.6 MeV, 2.3 MeV and 4.4 MeV 
varied inversely with solar activity in 
cycles 21--22 as expected from the theory of solar modulation of 
cosmic rays.  They were found to be strongly anticorrelated with cutoff
rigidity, as expected from the theory of the cutoff, falling by a factor
$\sim 3.6$ between the lowest ($< 7$ GV) and highest ($> 13$ GV) 
rigidities sampled.  The solar cycle modulation was particularly marked at
the lowest rigidities, reaching an amplitude of 16\%.

The ratios of the intensities of the lines produced by nuclear de-excitation 
(1.6 MeV, 2.3 MeV) and those from nuclear spallation (4.4 MeV) did not 
vary with either solar activity or cutoff rigidity, indicating that the 
shape of the
secondary neutron spectrum in the atmosphere above 5 MeV was approximately
constant over the times and regions sampled.  If it is
approximated by a power law in energy, we derive constraints on 
the absolute value of the power law index $\sim -1.15$ -- -1.45 and
better constraints on its
variability, $\le 5$\% over a solar cycle, and 
$\le 6$\% over {\em SMM\/}'s range of sampled cutoff rigidities.

We also measured 
the intensity of the electron-positron annihilation line at 0.511 MeV.  
This line also varies with the solar cycle, but its variation with 
cutoff rigidity is weaker than that of the nuclear lines,
falling by a factor 2 (rather than 3.6) over {\em SMM\/}'s 
range of sampled cutoff rigidities. 
This can be understood in terms of the energy dependences of the 
cross sections for positron production and for the hadronic interactions which
produce secondary neutrons.

\end{abstract}

\begin{article}

\section{Introduction}

A significant contribution to the radiation environment in low Earth 
orbit (LEO) consists of $\gamma$-radiation emitted by the atmosphere.  These 
so-called albedo photons are the ultimate products of impacts by high 
energy cosmic ray 
particles on the nuclei of atoms in the upper atmosphere.  Earth's 
magnetic field shields these layers to some extent from charged particles, 
especially the relatively soft-spectrum flux of protons from the Sun.  The 
more energetic particles in the Galactic cosmic radiation (GCR) are therefore 
responsible for the long-term quiescent atmospheric 
$\gamma$-radiation, away from times of discrete solar flaring and 
coronal mass ejection events.  

The intensity of the quiescent $\gamma$-ray flux
is expected to vary both spatially and temporally as a result of 
modulation of the GCR flux.  The latitudinal variation of the
geomagnetic field imposes a cutoff on the cosmic ray spectrum
corresponding to the vertical rigidity cutoff, while the 11-year
solar magnetic activity cycle modulates the overall incoming GCR
flux in an inverse sense.  

Early observations of Earth's $\gamma$-ray line spectrum were largely
made during balloon flights, which could not achieve wide spatial or
temporal coverage.  Ling [1976] discussed these measurements and presented
model source functions for lines independent of time and at fixed latitude.  
Two satellite missions provided much better coverage, and were much more
sensitive than the balloon experiments.  These were HEAO 3 during 1979--1980,
whose results were reported by Mahoney et al. [1981] and Willett and 
Mahoney [1992] (treating the strong 0.511 MeV electron-positron annihilation 
line, and six strong lines from nuclear interactions, respectively) 
and the {\it Solar Maximum Mission} ({\it SMM}), from which 
results are reported in this paper.

The Gamma Ray Spectrometer experiment (GRS) on board {\it SMM}
was active during 1980--1989 in LEO
and was heavily exposed to Earth atmosphere 
$\gamma$-radiation.  It sampled a broad range of the parameters which
affected this radiation; its orbital inclination of 28$^{\circ}$
intercepted geomagnetic cutoff rigidity values between 5--15 GV, and its 10-year
lifetime covered the second half of solar cycle 21 and the first half of cycle
22.  The HEAO 3 experiment was also in LEO, but at a higher inclination,
covering a wider range of cutoff rigidities; it had finer spatial and 
energy resolution than the GRS, but its lifetime was only 9
months and it was somewhat less sensitive than the GRS.

A spectrum accumulated by {\it SMM} during the first $3 \frac{1}{2}$
years of the mission at low cutoff rigidities was presented by
{\it Letaw et al.} [1989], and extended to cover the entire mission
by {\it Share and Murphy} [2001], by whom
some 20 broad and narrow lines were detected and
identified with known lines from high energy 
impacts upon atmospheric nitrogen and oxygen nuclei.  {\it Share and Murphy} 
[2001] also detected very broad residual features which could not
be so identified.  The strong line at 0.511 MeV due to the 
annihilation of positrons from GCR-induced electromagnetic air showers
was also detected.  {\it Letaw et al.} [1989] showed that the 
time-averaged line strength
measurements are in general agreement with the
values predicted by {\it Ling} [1976], and {\it Share and Murphy} found
agreement with the HEAO 3 measurements also [Willett and Mahoney 1992].

Our purpose here is to follow up this work by measuring the variation 
of the quiescent Earth atmosphere
$\gamma$-ray flux with time and latitude.  We will use as a surrogate
spatial variable the vertical geomagnetic cutoff rigidity, and likewise
temporal variability will be expressed in terms of the phase of the solar
cycle.

\section{Instrument, Observations and Analysis}

\subsection{Instrument and Observations}

The {\it SMM} orbit characteristics were: altitude $\sim$400--570
km, inclination 28$^{\circ}$, lifetime March 1980--December 1989.  
It was pointed at the Sun for almost the
whole of this time, and therefore observed Earth for an extended period
during each $\sim 96$-minute orbit.  The Earth disk subtended a radius
of about $70^{\circ}$ at this altitude, which was essentially contained
within the GRS's very large field of view (FOV).  
The accumulation of data was 
briefly interrupted (November 1983 -- April 1984) during 
preparations for repair of the satellite's attitude control.
The instruments were turned off
during passages of the South Atlantic Anomaly (SAA) to avoid damage from
high radiation doses from trapped particles.  Data taken within
$\sim 10^{4}$ s after SAA passages were not used because of intense
instrumental background $\gamma$-radiation from
short-lived radioactivities induced in the SAA.  The pattern of SAA
passages was strongly influenced by the precession period of the orbit,
which was about 48 d.

The spectrometer itself consisted of seven 7.6 $\times$ 7.6 cm NaI
detectors surrounded to the sides and rear by anticoincidence shielding
[{\it Forrest et al.} 1980].  It was sensitive to photon energies between 
0.3--8.5 MeV, with FWHM resolution $\sim 6$\% at 1 MeV and binning
into 476 energy channels.  The energy binning was kept consistent to very
good accuracy throughout the mission by active gain control using an on-board
$^{60}$Co calibration source.  The instrument response as a function of
energy and angle off-axis was calculated by Monte Carlo simulations
[{\it S. Matz} 1986, private communication].

The effective area of the detector at 1 MeV was 120 cm$^{2}$.
The FOV ($\sim 150^{\circ}$ at 1 MeV, FWHM, as in Figure 1{\em a\/} was
large enough to contain Earth's disk, as noted above.  The spectra were
accumulated over 1 min intervals and labeled according to "Earth angle"
(i.e. the angle between the Sun pointing and the center of Earth) and
geomagnetic cutoff rigidity.  They were screened to exclude 
$\gamma$-ray bursts, solar flares and geomagnetic disturbances.

Despite the strength of the Earth atmosphere spectrum, it was dominated
in the 1-minute spectra
by the background spectrum arising from radioactivity in the GRS
and spacecraft induced by energetic particle bombardment, mainly in
the SAA [{\it Share et al.} 1989].  This may
be eliminated by subtracting the count rates in spectra where the FOV
points away from Earth from those in spectra pointing towards Earth; the
spectra were selected to be within Earth angles 144$^{\circ}$--216$^{\circ}$
and $\le 72^{\circ}$, within the same orbit, and at similar
values of cutoff rigidity.  The rear shielding of
the detector was not sufficient to block Earth radiation completely
when pointing away from Earth, especially at high energies, so that when
the background subtraction was made the Earth spectrum partially canceled.
A correction factor for this effect, as a function of energy, was
derived by J. R. Letaw (unpublished report, 1988) from the Monte Carlo
simulations mentioned above, and was applied by us to the 
background-subtracted spectrum (Figure 1{\em b\/}).  
Otherwise, the background subtraction
was very effective, except for two residual lines at the $^{60}$Co
decay energies 1.17 and 1.33 MeV from the on-board source [{\it
Share and Murphy} 2001]. 

After background subtraction the spectra were summed over 3-day
intervals.  For studies of time and cutoff rigidity dependence the 3-day
spectra were summed into several combinations of time and rigidity
bins.  The basic unit for temporal studies was 48 days, in order to 
average over any effects arising from background variability on the
48 day precession time-scale.  In general the nuclear lines were found
to be too weak to be measured with good statistics in 48 days, so we
performed a further summation over 6 
months for temporal studies. We also made summations over 
9 years (i.e. the whole mission) for analyses of cutoff 
rigidity dependence. The rigidity bins generally employed 
were $<$7, 7--11 and $>$11 GV at 48 d and 6 month resolutions, and
$<$7, 7--9, 9--11, 11--13 and $>$13 GV over 9 years.  There is some
contamination of each rigidity bin by its neighbors, due to the GRS's
wide FOV, but this can be shown to be small, given the inverse square law
of flux and the width of the bins.  

A typical spectrum obtained in this way
is shown in Figure 2, where the lines identified by {\it Share and
Murphy} [2001] are plotted individually.  The unidentified very broad lines
detected by {\it Share and Murphy} [2001] are {\em not\/} plotted in Fig. 2.

A systematic uncertainty is expected in this analysis due
to $\gamma$-ray emission from astronomical sources, in particular the
Galactic center (GC).  In December of each year Earth's orbital motion
causes the apparent position of the Sun to pass very close in angle
to the GC, so that for some weeks around this time the GC was almost in the
center of the GRS FOV, until the Sun's motion (tracked by the GRS) carried 
it out of the FOV.  This emission was of course only visible when the
instrument pointed away from Earth; since these spectra were subtracted
from those pointing towards Earth (see above), the peak in emission due
to the GC's transit across the FOV ought to appear as a sharp dip
in the intensities of the components of our atmospheric 
spectra around the Decembers
of years 1980--1988 [{\it Share et al.} 1988].  In practice we found that
the dips were detected only in the time series of the narrow 0.511 MeV line,
the continuum, and the broad residual features mentioned in \S 1,
which are so broad that they must be contaminated by the continuum.  
Results for the 0.511 MeV line from the 6-month and 9-year spectra contained
unresolved dips, and were corrected by subtracting the averaged 
GRS exposure to the dip flux from the GC as a function of time 
(Fig. 10 of {\it Harris et al.} [1990]).  The correction factors were
small ($\sim 3--5$\%).

On the shorter 48-day time-scale, the dips are resolved, 
and are of interest in their own
right, since it is possible that an unknown terrestrial source might
exist, which exhibits a peak or trough every December.
We distinguished between celestial and terrestrial sources
of the 0.511 MeV line by looking at our
"Earth-viewing" and "sky-viewing" 
spectra separately on 48 day time-scales.  These are both contaminated 
by lines from radioactive decay, but the 0.511 MeV line from the GC is
strong enough to appear in the "sky-viewing" data as positive
annual peaks (instead of dips) due to the GC transits (Fig. 9
of {\it Share et al.} [1988]).  The "Earth-viewing" spectra should show
no signal from the GC (except for a small leakage through the rear of
the detector), responding only to terrestrial sources.

\subsection{Analysis Procedure}

We measured the strengths of the lines in spectra like those of 
Fig. 2 using a model 
spectrum containing 24 lines and the sum of two power-law continua between the
energies 0.65--8.5 MeV.  This was identical to the model used by {\it Share
and Murphy} [2001].  The lines were parametrized by 
energy, width and amplitude; the broad residual features mentioned in
\S 1 were modeled by 5 broad lines; the continua were parametrized by power 
law index and amplitude.  The model spectra were convolved
with the instrument response, with the parameters being varied until
the resulting predicted count spectra agreed with the observed spectrum,
as specified by the minimum value of the $\chi^{2}$ function.\footnote{ For
fits using such a large number of parameters the minima of the $\chi^{2}$
function may be shallow and ambiguous.  Our method of mapping the function
around the minimum is described by {\it Share and Murphy} [2001].  
In addition to the usual 
assignment of errors by the criterion of the minimum $\chi^{2}$ + 1, in
some fits systematic errors had to be included to take into account
the existence of more than one possible minimum.}  The model lines were 
assumed to be of Gaussian shape, and the power laws were constrained so 
that one dominated the continuum at low energies and the other at high
energies.

The positron annihilation line at 0.511 MeV is accompanied by a strong
continuum extending to lower energies, resulting from energy losses 
by Compton scattering in Earth's atmosphere.  We performed separate
fits to this feature, using a model consisting of a power law continuum
between 0.35--0.62 MeV,
a narrow line at 0.511 MeV, and a nearly flat continuum extending below
0.511 MeV to account for the scattered component, as illustrated in 
Figure 3.

\section{Results}

All of the lines detected by {\it Share and Murphy} [2001] could be
detected in our spectra, though with lower significance due to the
cuts we made into time periods and rigidity bins.  Both {\it Willett
and Mahoney} [1992] and {\it Share and Murphy} [2001] have made
detailed comparisons between {\it SMM} and HEAO 3 measurements.
Here we present results for those lines whose 
temporal or spatial behavior is expected to shed light on physical
processes in Earth's atmosphere.
  
Modelling indicates that the nuclear lines are mainly
due to interactions with secondary neutrons produced by GCR impacts, rather
than to the GCR themselves [{\it Ling} 1975, {\it Masarik and Beer} [1999].  
We may distinguish between lines produced
by de-excitation of excited states of N and O (which are
excited by neutrons of $\sim 6$--12 MeV), and those produced by
de-excitation of spallation products of N and O nuclei, which 
require higher energies (the cross section rises to a plateau above 
$\sim 20$ MeV).  The 4.44 MeV line is the main example of a
spallation-product de-excitation ({\it Ramaty, Kozlovsky and
Lingenfelter} [1979]  --- it arises from excited $^{11}$B or $^{12}$C, 
which are produced by spallation of $^{14}$N and $^{16}$O).  
Most of the other lines result from a combination of
spallation and de-excitation processes, as discussed in detail by
{\it Share and Murphy} [2001].  However the two lines at 1.635 and
2.311 in Fig. 2 are predominantly the results of de-excitation of 
$^{14}$N.  It follows that the ratio of the intensity of the 4.44 MeV
to the intensities of the 1.635 and 2.311 MeV lines (combined) is a 
diagnostic of the relative importance of spallation and de-excitation
reactions, and hence of the fluxes of secondary neutrons at high and
low energies respectively.  

We therefore present our results for, and focus our Discussion on, 
the three nuclear lines mentioned above and the 0.511 MeV positron
annihilation line. We discovered that, as expected, they and  
the continuum varied at least qualitatively with the
solar cycle, being strongest around solar minimum (mid-1986).  This is
illustrated for the strongest nuclear line in the spectrum, the 4.44 MeV
line from de-excitation of $^{12}$C and $^{11}$B,
in Figure 4{\em a\/}.  The phase of the solar
cycle is illustrated here by the atmospheric neutron fluxes measured at three
stations in the global neutron monitor system [{\it Shea and Smart} 2000]
whose geomagnetic cutoff rigidities correspond approximately to our rigidity
bins.\footnote{ The data were obtained from the archive at \\
http://www.env.sci.ibaraki.ac.jp/ftp/pub/WDCCR/STATIONS/. \\  
The stations were selected to have comparable altitudes, 
high count rates [Moraal et al. 2000], and to have been operating
during 1980--1989.  They are Alma Ata B, Kazakh SSR (rigidity 6.6 GV,
altitude 3340 m), Tsumeb, Namibia (rigidity 9.2 GV, altitude 1240 m), and 
Huancayo, Peru (rigidity 13 GV, altitude 3400 m).}  We quantify
the overall solar-cycle variability as the difference between
the average line fluxes during times of high activity (years 1984 through 
1988) and low activity (1980 through 1983, and 1989), which are presented 
in Table 1.

The figure and table also demonstrate that the 4.44 MeV
line intensity falls rapidly as cutoff rigidity increases; again, this is
expected due to the shielding from charged particles provided by
Earth's magnetic field.  The solar-cycle variation amplitude 
becomes weaker with
increasing cutoff rigidity; this is consistent with scaling as the weakening 
of the flux itself, but in the highest rigidity range ($>11$ GV) it is also
consistent with no variation.

The sum of the intensities of 
the 1.635 and 2.311 MeV lines is shown as a function of time 
in Fig. 4{\em b\/}.  It can be seen that
these lines also vary as expected with time (i.e. solar cycle)
and cutoff rigidity, as shown quantitatively in Table 1.  
The temporal behavior of the 0.511 MeV line is also
correlated with the solar cycle (Figure 5).  In this time series, 
which is at 48-day resolution, the 
dips around December due to subtraction of the annual transits of the
GC are clearly visible. The dips appear to be larger in the $< 7$ GV
cutoff rigidity data, a problem which will be addressed in \S 4.3.2. 
The upper envelope of the time series, 
if the dips are ignored, agrees qualitatively with the neutron monitor
solar cycle proxy.  The Compton and power law continua
exhibited a second set of strong annual dips due to emission from the
Crab Nebula, peaking in June of each year, which made it effectively
impossible to discern the shape of the upper envelope of their time
series.

To examine the rigidity dependence more fully, we give
our results in five rigidity bins from whole-mission spectra in Table 2, 
for the nuclear lines at 1.635, 2.311 and 4.44 MeV and for the
0.511 MeV annihilation line.  

\section{Discussion}

\subsection{The Behavior of the Lines from Spallation and Direct 
De-excitation}

As noted in \S 3, the ratio of the intensities of the lines at 4.44 MeV and
1.635 plus 2.311 MeV is a function of the relative importance
of spallation reactions and nuclear de-excitations caused by impacts of
neutrons on atmospheric nuclei.  This in turn depends on the relative
fluxes of secondary neutrons below 12 MeV and above 50 MeV respectively, 
due to the energy dependence of the cross sections, which we consider below.
In Figure 6, using the measurements in Fig. 4, we plot the ratio 
between the strengths of the 1.635 plus 2.311 MeV lines and the 4.44 MeV line
from 1980 to 1989. 
It is found to be essentially constant at all times and in all cutoff
rigidity bins --- in the figure, variability of the points within each
panel is not significant, nor are the differences between the long-term
averages of each panel.  The same result is also obtained from the results 
integrated over 9 years for the finer rigidity binning in Table 2 --- the
fluxes in both categories of line fall as cutoff rigidity increases, 
approximately in lockstep at a rate $\sim 15$\% per GV of rigidity.

We will compare these measured ratios of 1.635 plus
2.311 MeV to 4.44 MeV lines with theoretical predictions.  We generated the 
predictions from a neutron spectrum above 1 MeV at the top of the atmosphere 
(30 g cm$^{-2}$) at high rigidity (15 GV) calculated by {\it Kollar 
and Masarik} [1999] using Monte Carlo simulation.  We 
approximated this neutron spectrum by a broken power law \\
\indent $0.07 E^{-1.29}$ neutrons/(cm$^{2}$ s MeV)
~~~~~~for $E \le 10^{4}$ MeV \\
\indent $2.6 \times 10^{3} E^{-2.31}$
neutrons/(cm$^{2}$ s MeV) for $E > 10^{4}$ MeV  \\
\indent ~~~~~~~~~~~~~~~~~~~~~~~~~~~~~~(1) \\
We used the standard reaction rate formula [{\it Lang} 1980] in which this
distribution is folded with the cross sections for
neutron inelastic scattering and spallation on $^{14}$N.  These are available 
below 20 MeV from 
{\it Rogers et al.} [1975] --- above 20 MeV the inelastic cross sections go
to zero and the spallation cross section can be assumed to be constant at
its value $40 \pm 8$ mb at 20 MeV.  The 20\% uncertainty in the $^{14}$N
spallation cross section producing 4.44 MeV photons is the major source
of systematic error in this calculation.  Taking it into account,
we computed a predicted ratio of $0.36^{-0.06}_{+0.08}$, in very good 
agreement with our measurement of $0.39 \pm 0.03$ at high rigidity
$> 11$ GV (Table 1 and Fig. 6, right panel).  

We now quantify the extent to which the shape of this neutron spectrum is 
allowed to vary, given that our result is consistent with a shape varying
neither with time nor with cutoff rigidity (Fig. 6).  For this purpose
we take the low-energy power law index in Eq. (1) as the parameter of
interest, since the high-energy power law contributes little.  We repeated
our calculation of the (1.635 + 2.311 MeV)/4.44 MeV ratio with this index
allowed to vary, for $^{14}$N spallation cross section values in the
allowed range 32---48 mb.  The power law index was allowed to vary until
the $3 \sigma$ limits on the variation of the
line ratio from high to low phase of the solar cycle were exceeded (i.e.
$0.39 \pm 0.09$ at high cutoff rigidity $> 11$ GV).  Figure 7 shows the
results (left panel).  The dotted line represents the power law index
which reproduces the mean line ratio 0.39 as a function of the spallation
cross section and the full lines are the $3 \sigma$ limits.  It is clear
that, while the absolute value of the power law index may lie anywhere
within the range $\sim -1.2$--- -1.45, for a given $^{14}$N spallation
cross section it can only vary from the central value by about 5\% over time.

We next consider the allowable variation of the neutron spectral shape
with cutoff rigidity.  A further assumption about the shape is necessary
here, since predictions at cutoff rigidities in the 5--11 GV range
were not published by {\it Masarik and Beer}
[1999] and {\it Kollar and Masarik} [1999]; predictions were published for
high magnetic latitudes ($>70^{\circ}$ and gave a neutron spectrum 
shape similar to that at the equator.  We therefore make the very simple
assumption that the neutron energy spectrum at any rigidity has the same 
broken power law shape as in Eq. (1) such that the amplitudes of the two 
terms are multiplied by a common constant factor.  The calculation of the 
predicted (1.635 + 2.311 MeV)/4.44 MeV line ratio as a  
function of the low-energy power law index and the $^{14}$N spallation cross
section proceeded exactly as described above for the time variability
limits.  Our limit for the allowed variability of this ratio as a function
of cutoff rigidity was found from Table 2 to be $0.37 \pm 0.03$ (1 $\sigma$),
or $0.37 \pm 0.09$ (3 $\sigma$).  As would be expected from the constancy
of this ratio with both time and rigidity, the resulting limits on
the power law index are almost the same for spatial as for temporal
variability (Fig. 7 right): the absolute value may lie between 
$\sim -1.15$ --- -1.45, but
for any given $^{14}$N spallation cross section it can only vary by
about $\pm 6$\% between rigidities 5--15 GV.

In summary, despite the presence of a systematic uncertainty in its value
due to the experimental error in the nuclear cross sections,
the {\em variability\/} of the neutron spectrum's low-energy power law index
with time and cutoff rigidity is tightly constrained --- whatever 
the actual value of the index may be, the allowed changes during a
solar cycle, and over the range of cutoff rigidities 5--15 GV, 
are $\le 5$\%--6\% ($3 \sigma$).

\subsection{Amplitude of solar-cycle modulation}

It is quite clear from Figures 4{\em a\/} and 4{\em b\/} that in general
the nuclear $\gamma$-ray line fluxes are more strongly
modulated than the count rates from ground-based neutron monitors which
were selected to have similar geomagnetic cutoff rigidities.  This is
shown quantitatively
in Table 1 where we compare, for both the $\gamma$-ray lines and
the neutron monitors, the fluxes during low and high
cosmic ray activity periods.

At first sight this is unexpected, since the same incident cosmic ray spectra 
are producing the neutrons which are responsible for both measurements.
There is, however, an effect predicted by 
theoretical simulations of the neutron fluxes in the 
atmosphere, as a function of depth, which can explain this result.  
The key physical process is that that at the top of the atmosphere 
neutrons are able to escape upwards, 
and this leakage probability depends on the cosmic ray proton
energy spectrum; lower energy protons are more
likely to generate upward-leaking neutrons.  
Spacecraft of course see the effects of both upward and downward moving
neutrons, whereas ground based measurements see only the latter.  The
effect is illustrated in Figure 8
(after {\it Light et al.} [1973]).  Note that our results in \S 4.1
require that the {\em shape\/} of the neutron spectrum is practically
independent of the hardness of the proton spectrum --- it is the
{\em direction\/} of the neutrons (up or down) which is affected.
 \footnote{ This is illustrated in Fig. 8, where changes in the proton
energy have a very similar effect on
neutrons both above and below 10 MeV, (i.e. those causing both 
de-excitation lines and spallation lines).  This
is consistent with the leakage neutron spectrum remaining approximately
unchanged even when the incident proton spectrum changes substantially,
as required by our results in \S 4.1.}

A direct test of this solar modulation effect cannot be made, since no 
detailed simulations
have been published for the neutron spectrum as a function of
solar cycle phase and atmospheric depth.  However 
simulations as a function of cutoff rigidity and
depth were performed by {\it Kollar and Masarik} [1999], and the effect of
high rigidity is qualitatively similar to that of strong solar modulation,
i.e. a hardening of the incoming proton spectrum. In these simulations
the total neutron flux falls off much faster 
between the top and bottom of the atmosphere at low 
rigidities than at high rigidities.  Thus the neutron flux at a depth
$\sim 20$ g cm$^{-2}$ (as seen
from orbit) falls by a factor of 10 between cutoff rigidities 0.2 GV and
15 GV, while the neutron flux at the mountain-top level of the neutron 
monitors falls only by a factor 2.  The harder proton 
energy spectrum at 15 GV produces neutrons less likely to escape, and
therefore relatively more likely to reach 1000--3000 m.

This effect must also occur when changes in the proton 
energy spectrum arise
from the solar cycle modulation rather than from varying geomagnetic cutoff 
rigidity.  Thus the reduced amplitude of the modulation of the neutron monitor 
count rates relative to our line flux measurements is due to the softer
cosmic ray spectrum around solar minimum giving rise to relatively enhanced
leakage of neutrons which are visible to spacecraft but not to
ground-based instruments.  The softening of the spectrum is however much
less pronounced between the extremes of a solar cycle than between cutoff
rigidities 0.2--15 GV, so that the differences up to $\sim 8$\% between the 
modulations at the top and bottom of the atmosphere (Table 1) due to the
solar cycle are much less than the factor of 5 which the models predict 
to result from varying geomagnetic cutoff rigidity.

\subsection{The 0.511 MeV electron-positron annihilation line}

\subsubsection{Comparison with the nuclear de-excitation and spallation
lines}

The atmospheric 0.511 MeV positron
annihilation line owes its existence to pair 
production from the energetic photons generated in the electromagnetic 
component of air showers produced by very energetic GCR, not to 
secondary neutrons.  Although its intensity falls off with 
increasing cutoff rigidity $R$, Figure 9 (based on Table 2) clearly
show that it falls less rapidly than the intensity of the 4.44 MeV line.
The 4.44 MeV and other nuclear lines fall by factors of almost 4 when
cutoff rigidity increases from $< 7$ GV to $> 13$ GV, while the 0.511
MeV line falls off only by a factor $\sim 2$.

Some insight into the energy dependence of these primary processes can be
gained by comparing the dependence on $R$ of the lines with the
rate of fall-off of the downward cosmic ray proton flux with $R$, 
which was measured by the Alpha Magnetic Spectrometer (AMS) experiment
[{\it Alcaraz et al.} 2000] and is also plotted in Fig. 9.  The AMS
measurements covered proton energies $E_{p}$ in the range
70 MeV--200 GeV.  It is
seen that the 4.44 MeV line flux falls off approximately in step with the 
cosmic ray proton flux, whereas the 0.511 MeV line quite clearly falls
less steeply with increasing $R$.  The behavior of the 4.44 MeV line
implies that the multiplicity of the secondary neutrons 
which generate the 4.44 MeV line is not a strong function 
of $E_{p}$.\footnote{ As noted in \S 4.1, models suggest
that the {\em shape\/} of the secondary neutron spectrum is much less variable
with cutoff rigidity than the cosmic ray proton
spectrum [{\it Kollar and Masarik} 1999].  The $R$-dependence of the 4.44
MeV line might nevertheless differ from that of the proton flux if the 
multiplicity of neutrons produced per proton is a strong function of $E_{p}$.
The similarity between the curves for cosmic ray protons and the 4.44 MeV
line in Fig. 9 suggests that this is not the case.}
Model calculations of neutron production by nuclear
evaporation predict a neutron multiplicity $\sim E_{p}^{0.2}$ for N and O 
[{\it Konecn\'{a}, Valenta and Smutn\'{y}} 2000],
confirming that it is indeed a weak function of proton energy.

A second implication of the curves in Fig. 9 is that, conversely, the
air shower process which generates positrons and then 0.511 MeV 
photons is a strong function of energy --- the decreasing number of
cosmic ray protons as $R$ increases must be partly offset by the fact
that the remaining protons are more energetic and more efficient at 
producing positrons.  The electromagnetic cascade is produced by the
"soft" component of a charged particle induced shower.  Charged particle
strong interactions produce a shower of pions which may undergo 
further strong interactions (producing the "hard" component of the air 
shower), or decay into energetic $\gamma$-rays or leptons;
electron-positron pair production by electromagnetic interactions of these 
particles is well understood [{\it Lang} 1980].  The multiplicity
of secondary pions at these energies is a
rather weak function of energy, varying as
$E^{0.3}_{p}$: [{\it Wigmans} 2002]), but the cross section for pair 
production increases rapidly, as $\sim$ln$^{3}(E_{p})$, confirming our
expectation.
The relative $R$-dependences of the 4.44 and 0.511 MeV lines in Fig. 9
are therefore qualitatively explained by the $E_{p}$-dependences of the
multiplicities with which secondary particles (neutrons or pions) are
produced, and of the cross sections of the reactions which follow.

\subsubsection{Anomalous rigidity dependence of the Galactic 0.511 MeV line}

The annual dips in the time series of 0.511 MeV line fluxes, which are
clearly visible in Fig. 5, are apparently dependent upon
$R$: the dips in the bin $<7$ GV cutoff rigidity are 
about twice as deep as those in the other bins.  This is not
expected from the theory of GC transits developed by {\it Share et al.}
[1988] since (given the constant GC source) the dip 
amplitude depends only on the occultation geometry, which is not expected to
vary significantly with rigidity.  There are three possible causes:
selection effects
in the transit geometry at low rigidity, the transiting of a second,
unknown cosmic 0.511 MeV line source across the GRS FOV, or 
an unexpected effect intrinsic
to the LEO environment which happens to cause a
drop in the atmospheric 0.511 MeV line flux every December.  

We tested for the presence of a cosmic or terrestrial source of the dips by
dividing our spectra into "Earth-viewing" and "sky-viewing" periods as 
described in \S 2.1; the dips in a time series of "Earth-viewing minus 
sky-viewing" 0.511 MeV amplitudes should appear as peaks in the "sky-viewing"
subset, and not at all in the "Earth-viewing" subset.  The 
$R$ dependence of the dips in Fig. 5 might be caused by a rigidity
dependence of the "sky-viewing" peaks (as in the first two hypotheses above)
or by rigidity-dependent dips in the "Earth-viewing" subset (as in the third
hypothesis).  We fitted 48-day
accumulated spectra of each type between 0.35--0.75 MeV 
(sub-divided into subsets with $R > 11$
and $R < 11$) with a model spectrum containing four
known lines from spacecraft radioactvity and the 0.511 MeV, plus a power law
continuum (there is a full description in {\it Share et al.} [1988]).
The resulting amplitudes for the 0.511 MeV line are shown in Figures
10{\em a\/} and 10{\em b\/}.
It is clear that the annual "sky-viewing" peaks are not significantly
dependent upon $R$.  The "Earth-viewing" spectra, by contrast, show dips
which are much stronger at low rigidity.  However, the relation between 
these dips and December of the calendar is not very strong.

This analysis shows that the effect is terrestrial in origin.\footnote{ We 
confirmed that the source is not celestial by
calculating the exposures of the GRS to the GC and other possible cosmic 
sources when in the $R < 7$ and $R > 11$ GV portions of the orbit (cf. {\it
Share et al.} [1988], {\it Harris et al.} [1990]).  We found no significant
differences.}  The apparent strengthening of the GC-induced dips in
December in Fig. 5 appears to be a chance occurrence of a small number
of $R$-dependent dips at that time of year (Fig. 10{\em b\/}).

There is no immediately obvious candidate for an effect intrinsic to 
Earth's atmosphere which
is restricted to low geomagnetic cutoff rigidities and operates 
sporadically on a 48-day time-scale.  The dips do not appear to be more
frequent at any phase of the solar cycle, or to correlate with
geomagnetic events.  Equally, an
artefact of unknown origin in the instrument or in our analysis may be
responsible.

\subsubsection{Comparison with HEAO 3 results}

The 0.511 MeV annihilation line was strong enough for the HEAO 3 count rates
at this energy to be broken down into bins of geomagnetic cutoff rigidity,
much as we have done (\S 2.1).  We compare our result (Table 2) with that of 
{\it Mahoney, Ling and Jacobson} [1981] in Fig. 9 (open versus closed
circles).  The HEAO 3 flux values per steradian have been integrated over
the {\it SMM} exposure to the $68^{\circ}$ Earth disk.
This integration was also weighted by the limb darkening function which
{\it Mahoney, Ling and Jacobson} [1981] measured using HEAO 3's superior
spatial resolution, which is proportional to $1 + 1.7$ cos$\theta$ 
where $\theta$ is the satellite zenith angle seen from a disk 
point. 
The HEAO 3 measurements then agree well with the {\it SMM} values,
though having a rather
shallower decrease with cutoff rigidity.  They are consistent with our finding
that the 0.511 MeV line falls off more slowly with rigidity than the 4.44 MeV
line, and with our explanation in terms of the interaction cross sections
involved (\S 4.2).

\subsection{Broad residual lines at 1--5 MeV}

We noted in Fig. 2 that, in addition to the narrow lines shown there, 
five broad lines were necessary to fit the spectrum [{\it Share and
Murphy} 2001].  The actual residual features which are fitted by these
five lines are shown in Figure 11.  The peak around 5 MeV suggested to
{\it Share and Murphy} [2001] that thermal or epithermal neutron capture on the
$^{127}$I in the NaI and CsI detectors was responsible.  
We repeated their comparison between the I thermal neutron $\gamma$-ray
spectrum and the residual features using an updated database of I lines.
Figure 11 shows that the agreement is worse than that found by
{\it Share and Murphy} [2001], although the discrepancy at energies
$< 1$ MeV when normalized to the 5 MeV feature is only a factor $\sim 2$.  

Our result in Fig. 11 suggests that the residual feature in the 1--3
MeV range is due mainly to a different process from the instrumental 
neutron captures which generate the $\sim 5 $ MeV peak.  We propose that
there are atmospheric nuclear lines in this range which are not resolved by 
the GRS.  Additional evidence 
for this comes from the dependences of the strengths of 
the 1--3 and 5 MeV residual features on geomagnetic cutoff rigidity in the 
range $R \le 13$ GV (Figure 12).  In \S 4.1 we noted that 
atmospheric nuclear lines in general
decrease at approximately the same rate when rigidity increases --- the
characteristic behavior, taking the 4.44 line as typical, is a decrease
in strength by about 15\% per GV (Table 2).  The broad 1--3 MeV residual
feature shares this behavior: the ratio between this feature
and the 4.44 MeV line is plotted in Figure 12 and seen to be approximately
constant below 13 GV --- there is clearly some other process at work above 13
GV (see below).  The constancy below 13 GV is consistent with the 
feature being composed of 
unresolved atmospheric lines from nuclear spallation reactions or nuclear 
de-excitation (the underlying reason being the constancy of the secondary
neutron spectrum, \S 4.1).

Compared with the 15\% per GV rigidity fall-off rate of both the
atmospheric nuclear lines and the 1--3 MeV feature,
the 5 MeV residual feature in Fig. 12 falls off 
much faster for $R \le 13$ GV.  This is expected if it arises from
thermal or epithermal neutron captures in orbit, i.e. the upward
leakage neutrons whose behavior was discussed
in \S 4.2.  In our comparison of neutron monitor and satellite data we 
found that as $R$ increases the primary GCR proton spectrum
becomes much harder, whereas upward-leaking neutrons are more efficiently
produced by low energy protons (Fig. 8).  Just as relatively
more neutrons reach ground-based neutron monitors at high $R$, relatively
fewer neutrons will reach LEO --- i.e. the ratio between 
LEO neutrons and atmospheric neutrons gets smaller as rigidity increases,
as does the ratio between the 1--3 MeV and $\sim 5$ MeV features in Fig. 12
below 13 GV.

It is clear from Fig. 12 that at the highest rigidities $R > 13$ GV (close to
the geomagnetic equator) there is a different process acting.  The 5
MeV residual feature rises sharply, and the 1--3 MeV residual feature
shows a less marked increase, relative to the 4.4 MeV line.  We suggest
that these increases result from the same epithermal neutron captures on
I, but that there is a distinct equatorial population of neutrons over and
above the up-going neutrons produced directly by GCR (whose abundance 
is lowest at
the equator).  One possible source of such neutrons is the LAP (low altitude 
particles) detected at energies above 500 keV by
{\it Hovestadt et al.} [1972] at low geomagnetic latitudes.  These ions are
ultimately derived from the equatorial ring current, but extend much lower
in altitude (down to $< 100$ km).  Their abundance depends only weakly
on solar activity [{\it Mazur, Mason and Greenspan} 1998].  
Their energy spectrum 
extends at least up to 35 MeV [{\it Gusev et al.} 1996], well above the 
$\sim 6$ MeV thresholds of the reactions 
$^{14}$N($p$,$n$)$^{14}$O and $^{14}$N($\alpha$,$n$)$^{17}$F.  Neutrons 
produced in this way would have low (epithermal) energies, as required for
the 5 MeV residual feature, making a small contribution to the 1--3 MeV
feature also (Fig. 11, dashed line).  
A weak point of this hypothesis is that the spectrum measured by {\it Mizera
and Blake} [1973] and
{\it Gusev et al.} [1996] is very soft (varying as energy $E^{-4.4}$ if a
power law), so that relatively few ions exceed the reaction thresholds.

Alternatively, albedo charged particles 
(electrons --- {\it Basilova et al.} [1979]; protons --- {\it Efimov et al.}
[1985], referred to by {\it Alcaraz et al.} [2000] as the 
"second spectrum") are known to rise in numbers near the geomagnetic equator.
Unlike the albedo neutron population, their fluxes are not related in a simple
way to the primary GCR fluxes, due to their very complex trajectories in the
geomagnetic field.  Like the neutrons, they are thought to originate in 
GCR impacts on atmospheric nuclei.  The albedo protons have
moderately hard spectra extending up to a few GeV and thus occur above 
the threshold energy for neutron production.  A drawback to this model is
the uncertainty in the propagation of the particles in the magnetic field,
which was derived by {\it Alcaraz et al.} [2000] from Monte Carlo simulations
rather than observation.  The simulations suggest that only a subset of
albedo protons --- those whose trajectories involve many mirrorings over
a relatively long lifetime ("quasi-trapped") --- are equatorially 
concentrated, and it is
suggested that their origin is associated with the energetic trapped SAA
particles.  The flux is not expected to be isotropic, having a strong
east-west gradient, which makes the directionality of any resultant neutrons,
and {\em SMM\/}'s response to them, very difficult to predict.  The
properties of these particles are reviewed by Huang [2001].

\section{Summary}

We suggest four conclusions as to the behavior of the atmospheric 
$\gamma$-ray lines
which we have studied here.  \\
(1) The temporal behavior of all lines
can be explained by the varying modulation of the incident GCR flux during
the solar cycle.  \\
(2) The lines produced by two
processes arising from secondary neutrons (spallation and de-excitation
of air nuclei) behave similarly with respect to
geomagnetic location (i.e. cutoff rigidity).  \\
(3) It follows from this that the shape of secondary neutron energy 
spectrum is not strongly variable over the 
solar cycle or over the limited range of cutoff rigidities sampled by 
{\it SMM}.  If we assume a simplified broken power-law spectrum of 
secondary neutrons, then the low energy power law index must
lie in the approximate range -1.15 -- -1.45. There is a systematic error in
the absolute value, but the variability in that absolute value over a solar
cycle must be $< 5$\%; the spatial variability over the 
cutoff rigidity range 5--15 GV must likewise be $< 6$\%
(both $3 \sigma$ upper limits). \\
(4) The behaviors of these neutron-induced lines and of 
the 0.511 MeV positron annihilation line with respect to 
cutoff rigidity are different, and the difference
can be explained by the rigidity dependence of the measured GCR proton flux
and by the cross sections for the elementary particle
interactions which produce them (respectively,
secondary neutron production, and electromagnetic cascades from
energetic secondary pions).

We find that the signal from the known Galactic 0.511 MeV line (superposed on
that from Earth's atmosphere) unexpectedly shows a dependence on cutoff
rigidity (\S 4.3.2), which appears to be due to the infrequent chance 
occurrence of unexplained short decreases in the Earth-atmosphere 0.511 MeV
line signal at higher latitudes.

The origins of two very broad residual features in the spectrum 
($\sim 1$--3 MeV and 5 MeV, 
which we modeled by five broad Gaussian lines) appear to be complex.  We
tentatively suggest that most of the 1--3 MeV feature comes from the
contribution of many weak unresolved atmospheric lines, while the rest of
the 1--3 MeV feature and the whole of the 5 MeV feature are due to
the capture of ambient thermal or epithermal neutrons in the iodine in the GRS
detectors.  If true, this requires the existence of two distinct populations
of neutrons in LEO, one of which follows the cutoff rigidity law described
above (second and third conclusions) while the other is confined to low
geomagnetic latitudes.

\begin{acknowledgements}
We are grateful to Ben-Zion Kozlovsky of Tel Aviv University and Ron Murphy
of NRL for helpful discussions, and to two referees for their suggested
improvements.  This work was supported by NASA grant DPR W19916. 
\end{acknowledgements}

\end{article}

\newpage

\begin{figure}

\caption{(a) Effective area of the GRS as a function of distance of source 
off-axis, relative to on-axis ($0^{\circ}$), at 1 MeV.  (b) Correction
factor for the GRS response to a source subtending $68^{\circ}$ (as
Earth from 500 km altitude) due to leakage of $\gamma$-rays through the
rear of the detector when pointed away from Earth, {\it Letaw} [1988])}

\caption{Count spectrum observed by the {\em SMM\/} GRS 
from Earth's atmosphere accumulated in the
geomagnetic cutoff rigidity range 7--9 GV during 1980--1989.  Data points
--- measurements.  Thin full lines --- the instrument's response to 
individual lines identified by 
{\it Share and Murphy} [2000] superimposed on a two-power-law continuum 
(lower envelope of lines).  Five very broad line-like features of unknown 
origin are not plotted, either individually or combined with the
continuum, for the purpose of clarity (see Fig. 11).  The
sum of all the components as fitted to the measured spectrum is also shown
(thick full line).  The positions of the important lines at 1.635, 2.311
and 4.44 MeV are shown by arrows.}

\caption{Count spectrum observed by the {\em SMM\/} GRS from Earth's
atmosphere accumulated at cutoff rigidities $> 11$ GV during the 48 d
interval 1986 December 6---1987 January 19, compared with a fitted model
spectrum.  Data points---measurements.  Dotted line---power law
continuum component.  Dashed line---Compton scattered 
component arising from 0.511 MeV line.  Dot-dashed line---0.511 MeV line
component.  Full line---sum of three components.}

\caption{(a) Flux measured in the 4.44 MeV line 
in six-month periods during 1980--1989 in intervals of geomagnetic cutoff
rigidity $<7$ GV (top), 7--11 GV (middle), $>11$ GV (bottom). (b)
Sums of measurements of the fluxes of the 1.635 and 2.313 MeV lines in the
same six-month periods and rigidity intervals.  Dashed lines --- 
neutron monitor measurements of atmospheric neutron
abundance at Alma Ata B (top), Tsumeb (middle), and Huancayo (bottom)
normalized to the line fluxes by simple multiplication (see footnote 2).  All
fluxes are corrected for detector aperture and efficiency.}

\caption{Data points --- Flux measured in the 0.511 MeV line 
in 48-day periods during 1980--1989 in intervals of geomagnetic cutoff
rigidity $<7$ GV (top), 9--11 GV (middle), $>13$ GV (bottom).
Full lines --- neutron monitor measurements from Alma Ata B (top),
Tsumeb (middle) and Huancayo (bottom),
normalized to the line fluxes by simple multiplication (see footnote 2).
All fluxes are corrected for detector aperture and efficiency.}

\caption{Data points --- Ratio of the sum of the six-monthly 1.635 
and 2.313 MeV line strengths to that of the 4.44 MeV line (see Fig. 4) 
for cutoff rigidity intervals $<7$ GV ({\it left}), 7--11 GV 
({\it center}), and $>11$ GV ({\it right}).
Full lines --- mean value of the ratio during the mission for each 
rigidity interval.  Dash lines --- $1 \sigma$ errors on the mean values.}

\newpage

\caption{Allowed range of variability of the energy spectrum of 
atmospheric neutrons at 30 g cm$^{-2}$ depth, in the approximation
of two power laws in energy broken at $10^{4}$ MeV.  Variability is
parametrized by the values of the low energy power law index permitted
by {\em SMM\/}'s measured limits on the $\gamma$-ray line ratio
(1.6 + 2.3)/4.4 MeV in Fig. 6.  The effect of the experimental uncertainty
in the $^{14}$N($p$,$x$)$^{12}$C$^{*}$(4.44 MeV) cross section at 20 MeV
is shown (abscissa).  {\em Left panel\/} --- variability in time allowed by the 
{\em SMM\/} measurement of the line ratio at rigidities $> 11$ GV
$0.39 \pm 0.03$ (see Fig. 6 right panel).  {\em Right panel\/} --- variability
with cutoff rigidity allowed by the {\em SMM\/} measurement 
$0.37 \pm 0.03$ (from range of mean values
of all three panels in Fig. 6).  In both panels the dotted line is the power
law index which reproduces the measured line ratio 
for a given $^{14}$N + $p$ cross section and the solid lines 
are the limits which are consistent with that ratio at $3 \sigma$.}

\caption{Column-integrated rates of neutron production
by monoenergetic protons incident at the top of the atmosphere, 
and of neutron upward leakage, computed
by {\it Light et al.} [1973].  The fraction of leakage neutrons below
10 MeV is also shown.}

\caption{Intensities of the 0.511 MeV line ({\it open circles}) and the 
4.4 MeV line ({\it crosses}) measured from spectra accumulated in five 
geomagnetic cutoff rigidity intervals during the entire mission (Table
2).  Filled circles --- intensities of the 0.511 MeV line measured by
the HEAO-3 spectrometer [{\it Mahoney, Ling and Jacobson} 1981].
Full line --- downward flux of cosmic ray protons with energies
between 0.07--200 GeV measured by AMS.  The 4.4 MeV and proton fluxes are 
normalized to agree with the {\it SMM} 0.511 MeV line flux for rigidities $<7$
GV (Table 2, footnote 4). The HEAO-3 measurements are corrected to 
account for the GRS FOV and for limb darkening as described in the text.}

\caption{(a) Flux measured in the 0.511 MeV line in 48-day periods at cutoff
rigidities $>11$ GV and $<11$ GV when the GRS was pointed away from Earth
("sky-viewing"; Earth angles $144^{\circ}--216^{\circ}$).  Note the long-term
trend due to buildup of $\beta^{+}$-emitting radioactivities, with
annual peaks superimposed due to transits of the GC across the aperture.
(b) Flux measured in the 0.511 MeV line in 48-day periods at cutoff
rigidities $>11$ GV and $<11$ GV when the GRS was pointed towards Earth
(Earth angles $<72^{\circ}$).  Carets show the epochs of GC transits.}

\caption{The count spectrum of Fig. 2 with the continuum and narrow lines
subtracted.  Data points --- the residual spectrum after the 
subtraction.  Full line --- the fitted sum of all spectral features as 
in Fig. 2.  Dotted lines --- five broad line-like features not shown in
Fig. 2, which fit the residual spectrum.
Dashed line --- the response of the {\em SMM\/} GRS to
the line spectrum from thermal neutron capture on iodine, normalized to
agree with the broad feature at 5 MeV.}

\caption{Ratio of the count rates in the broad features 1--3 MeV (full
line) and 5 MeV (dashed line) to the strength of the 4.44 MeV line as
a function of cutoff rigidity.  The lower bound of the 5 MeV feature is 
assumed to be 3.8 MeV, and its ratio to 4.44 MeV is multiplied by 8 for
clarity.}

\end{figure}

%
%
%
%

\clearpage

\begin{table}
\begin{center}
\begin{tabular}{lclccc}
\tableline
Line or  & & & Low activity & High activity & Ratio \\ 
quantity & Spectrum & Interval  & average flux\tablenotemark{1} & average 
flux\tablenotemark{1} & High: \\
averaged & (rigidity) & & (1980--1983, 1989) & (1984--1988) & Low  \\
\tableline
4.44 MeV & $<7$ GV & 6 month & $11.1 \pm 0.2$ & $12.8 \pm 0.2$ &
$1.16 \pm 0.03$  \\
 & 7--11 GV & 6 month & $6.4 \pm 0.1$ & $7.4 \pm 0.2$ & $1.14 \pm 0.03$  \\
 & $>11$ GV & 6 month & $3.9 \pm 0.1$ & $4.1 \pm 0.1$ & $1.03 \pm 0.03$ \\
1.6 + 2.3 MeV & $<7$ GV & 6 month & $4.2 \pm 0.1$ & $4.9 \pm 0.1$ & $1.16 \pm
0.05$ \\
 & 7--11 GV & 6 month & $2.4 \pm 0.1$ & $2.6 \pm 0.1$ & $1.09 \pm 0.05$ \\
 & $>11$ GV & 6 month & $1.6 \pm 0.1$ & $1.5 \pm 0.1$ & $0.92 \pm 0.09$ \\
0.511 MeV & $<7$ GV & 48 day & $41.3 \pm 0.1$ & $ 45.7 \pm 0.1$ & 
$1.104 \pm 0.004$ \\
 & 7--11 GV & 48 day & $29.9 \pm 0.1$ & $31.6 \pm 0.1$ & $1.056 \pm 0.004$ \\
 & $>11$ GV & 48 day & $23.3 \pm 0.1$ & $22.2 \pm 0.1$ & $1.051 \pm 0.005$ \\
Alma Ata B neutrons & 6.6 GV & 6 month & 8709 & 9385 & 1.08 \\
Tsumeb neutrons & 9.2 GV & 6 month & 11290 & 11865 & 1.05 \\
Huancayo neutrons & 13 GV & 6 month & 1711 & 1760 & 1.03 \\
\tableline
\\

\end{tabular}
\end{center}

\tablenotetext{1}{Gamma-ray line fluxes in units of $10^{-3}$ photon 
cm$^{-2}$ s$^{-1}$.  The low and high cosmic ray activity 
periods were defined by the Alma Ata B 
count rates being either above or below the mean for the
period 1980--1989.}
\caption{Comparison of average fluxes in selected $\gamma$-ray lines during
periods of high and low solar activity as defined by the Alma Ata B neutron
monitor (cf. Figs. 4 and 5).}

\end{table}

\clearpage

\begin{table}
\begin{center}
\begin{tabular}{cccccc}
\tableline
Rigidity & Live time & Sum flux & Flux & Flux\tablenotemark{1} & Downward  \\
cutoff & seconds & 1.635 MeV & 4.44 MeV & 0.511 MeV & cosmic ray proton \\
GV &  & + 2.311 MeV &  &  & flux \\
\tableline
$<$7 & $3.64 \times 10^{6}$ & 1\tablenotemark{2} &  1\tablenotemark{3}
 & 1\tablenotemark{4}  & 1\tablenotemark{5} \\
7--9 & $2.95 \times 10^{6}$ & $0.66 \pm 0.04$ & $ 0.73 \pm 0.02$ & $0.805 \pm 
0.007$ & 0.606 \\
9--11 & $4.08 \times 10^{6}$ & $0.51 \pm 0.04$ & $0.49 \pm 0.01$ & $0.624 \pm 
0.005$ & 0.444 \\
11--13 & $5.89 \times 10^{6}$ & $0.34 \pm 0.02$ & $0.37 \pm 0.01$ & $0.532 \pm 
0.005$ & 0.348 \\
$>13$ & $1.25 \times 10^{6}$ & $0.27 \pm 0.03$ & $0.28 \pm 0.01$ & $0.479 \pm
0.005$ & 0.321 \\
\tableline
\\

\end{tabular}
\end{center}

\tablenotetext{1}{Corrected for contamination by the GC 0.511 MeV line
source as described in the text, footnote 3.}
\tablenotetext{2}{Value before normalization: $4.5 \pm 0.1 \times 10^{-3}$
photon cm$^{-2}$ s$^{-1}$.}
\tablenotetext{3}{Value before normalization: $1.19 \pm 0.02 \times 10^{-2}$
photon cm$^{-2}$ s$^{-1}$.}
\tablenotetext{4}{Value before normalization: $4.38 \pm 0.03 \times 10^{-2}$
photon cm$^{-2}$ s$^{-1}$.}
\tablenotetext{5}{Value before normalization: 412.7 proton m$^{-2}$ s$^{-1}$
sr$^{-1}$.}
\caption{Fluxes in selected $\gamma$-ray lines during the entire mission 1980--1989 as a function of geomagnetic vertical cutoff rigidity, normalized to rigidity $< 7$ GV.}

\end{table}


\begin{thebibliography}{}

\bibitem{}
Alcaraz, J., et al., Protons in near Earth orbit, $Phys. 
Lett. B$, $472$, 215--226, 2000.
\bibitem{}
Basilova, R. N., E. I. Kogan-Laskina, G. I. Pugacheva, and I. A. Savenko,
The quasi-trapped electron flux with E$>80$ MeV at an altitude of 200--250
km, in $Proc. 16th Int. Cosmic Ray Conf.$ (Kyoto), $3$, 150--153, 1979.
\bibitem{}
Efimov, Yu. E., A. A. Gusev, K. Kudela, L. Just, and G. I. Pugacheva, Spatial
distribution of albedo particles on altitudes $\sim$500 km, $Czechoslovak J.
Phys.$, $B35$, 1371--1381, 1985.
\bibitem{}
Forrest, D. J., et al., The gamma ray spectrometer for the
Solar Maximum Mission, $Solar Phys.$, 65, 15--23, 1980.
\bibitem{}
Gusev, A. A., T. Kohno, W. N. Spjeldvik, I. M. Martin, G. I. Pugacheva, 
and A. Turtelli Jr., Dynamics of the low-altitude energetic proton fluxes
beneath the main terrestrial radiation belts, $J. Geophys. Res.$, $101$, 
19659--19663, 1996.
\bibitem{}
Harris, M. J., G. H. Share, M. D. Leising, R. L. Kinzer, and D. C. Messina,
Measurement of the 0.3--8.5 MeV Galactic gamma-ray spectrum from the
Galactic center direction, $Astrophys. J.$, $362$, 135--146, 1990.
\bibitem{}
Hovestadt, D., B. H\"{a}usler, and M. Scholer, Observation of energetic
particles at very low altitudes near the geomagnetic equator, $Phys. Rev.
Lett.$, $28$, 1340--1344, 1972.
\bibitem{}
Huang, M. A., Physics results from the Alpha Magnetic Spectrometer 1998
Shuttle flight, in $Proc. 7th Taiwan Astrophysics Workshop$, ed. C-M.
Ko, World Scientific, Singapore, 2001.  LANL preprint $astro-ph$/0104229.
\bibitem{}
Kollar, D., and J. Masarik, Numerical simulation of particle fluxes and
cosmogenic nuclide production rates in Earth atmosphere, $Acta Physica
Universitatis Comenianae$, $40$, 81--93, 1999. \\
http://javier.dnp.fmph.uniba.sk/$\sim$dkollar/work/acta1999.pdf, 1999.
\bibitem{}
Konecna, \'{A}., V. Valenta, and V. Smutn\'{y}, Calculations of spallation 
neutrons yields from targets, $Third International Conference on ADTT and
Applications$, Paper P-e26, Prague, June 7--11 1999. \\
http://www.fjfi.cvut.cz/Stara\_verze/k417/web\_ads/papers/P-e26.pdf, 2000.
\bibitem{}
Lang, K. R., $Astrophysical Formulae$, Springer-Verlag,
Berlin, 1980.
\bibitem{}
Letaw, J. R., G. H. Share, R. L. Kinzer, R. Silberberg, E. L. 
Chupp, D. J. Forrest, and E. Rieger, Satellite obsevations of atmospheric
nuclear gamma radiation, $J. Geophys. Res.$, $94$, 1211--1221, 1989.
\bibitem{}
Light, E. S., M. Merker, H. J. Verschell, R. B. Mendell and S. A. Korff,
Time dependent worldwide distribution of atmospheric neutrons and of their
products, $J. Geophys. Res.$, $78$, 2741--2762, 1973.
\bibitem{}
Ling, J. C., A semiempirical model for atmospheric $\gamma$
rays from 0.3 to 10 MeV at $\lambda = 40^{\circ}$, $J. Geophys. Res.$,
$80$, 3241--3252, 1975.
\bibitem{}
Mahoney, W. A., J. C. Ling, and A. S. Jacobson, HEAO 3 measurements of
the atmospheric positron annihilation line, $J. Geophys. Res.$, $86$,
11098--11104, 1981.
\bibitem{}
Masarik, J., and J. Beer, Simulation of particle fluxes and cosmogenic
nuclide production in the Earth's atmosphere, $J. Geophys. Res.$, $104$,
12099--12111, 1999.
\bibitem{}
Mazur, J. E., G. M. Mason, and M. E. Greenspan, The elemental composition
of low altitude 0.49 MeV/nucleon trapped equatorial ions, $Geophys. Res.
Lett.$, $25$, 849--852, 1998.
\bibitem{}
Mizera, P. F., and J. B. Blake, Observations of ring current protons at low
altitudes, $J. Geophys. Res.$, $78$, 1058--1062, 1973. 
\bibitem{}
Moraal, H., A. Belov, and J. M. Clem, Design and co-ordination of 
multi-station international neutron monitor networks, 
$Space Sci. Rev.$, $93$, 285--303, 2000.
\bibitem{}
Ramaty, R., B. Kozlovsky, and R. E. Lingenfelter, Nuclear
gamma-rays from energetic particle interactions, $Astrophys. J. Suppl.
Ser.$, $40$, 487--526, 1979.
\bibitem{}
Rogers, V. C., V. J. Orphan, C. G. Hoot, and V. V. Verbinski, Gamma-ray
production cross sections for carbon and nitrogen from threshold to 20.7 MeV,
$Nucl. Sci. Eng.$, $58$, 298--313, 1975.
\bibitem{} 
Share, G. H., R. L. Kinzer, J. D. Kurfess, D. C. Messina, W. R. Purcell,
E. L. Chupp, D. J. Forrest, and C. Reppin, SMM detection of diffuse
Galactic 511 keV annihilation radiation, $Astrophys. J.$, $326$, 717--732, 1988.
\bibitem{}
Share, G. H., R. L. Kinzer, M. S. Strickman, J. R. Letaw, E. L. Chupp, D. J.
Forrest, and E. Rieger, Instrumental and atmospheric background lines observed
by the SMM Gamma Ray Spectrometer, in $High Energy Radiation Background in
Space$ (AIP Proceedings 186), ed. A. C. Rester, Jr., and J. I. 
Trombka, (New York: AIP), 266--277, 1989.
\bibitem{}
Share, G. H., and R. J. Murphy, Atmospheric gamma rays from
solar energetic particles and cosmic rays penetrating the magnetosphere,
$J. Geophys. Res.$, $106$, 77--92, 2001.
\bibitem{}
Shea, M. A., and D. F. Smart, Fifty years of cosmic radiation data,
$Space Sci. Rev.$, $93$, 229--262, 2000.
\bibitem{}
Wigmans, R., PeV cosmic rays : a window on the leptonic era?,
$Astroparticle Physics$, $19$, 379--392, 2003.
\bibitem{}
Willett, J. B., and W. A. Mahoney, High spectral resolution 
measurement of gamma ray lines from the Earth's atmosphere, 
$J. Geophys. Res.$, $97$, 131--139, 1992.


\end{thebibliography}
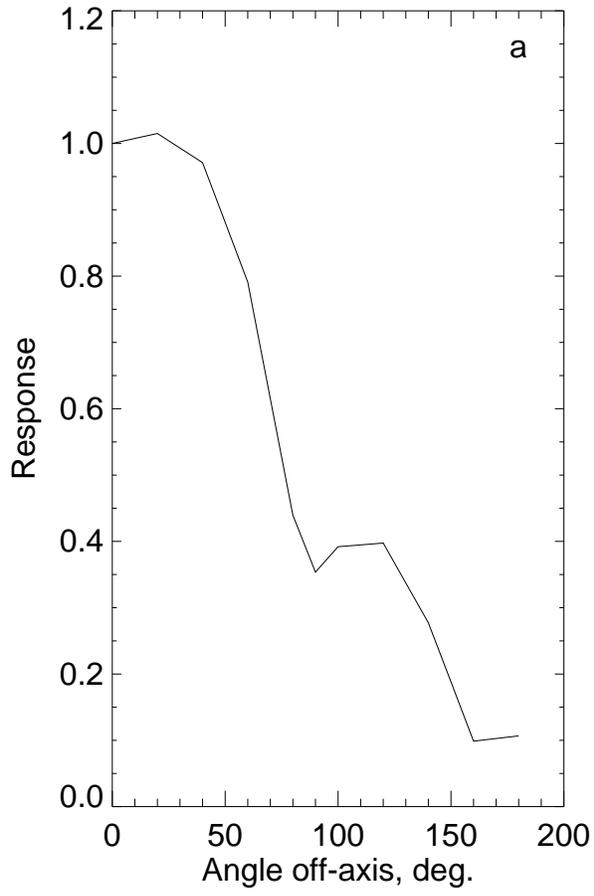
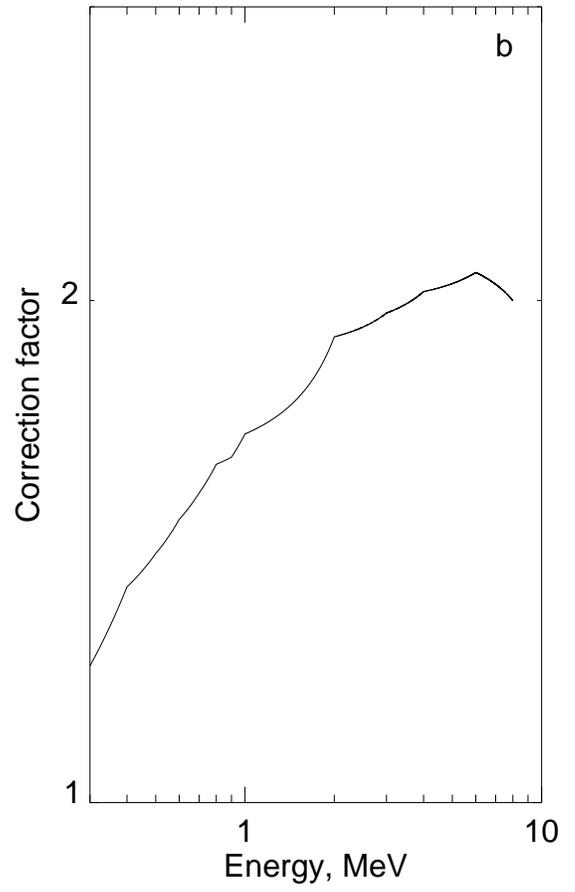
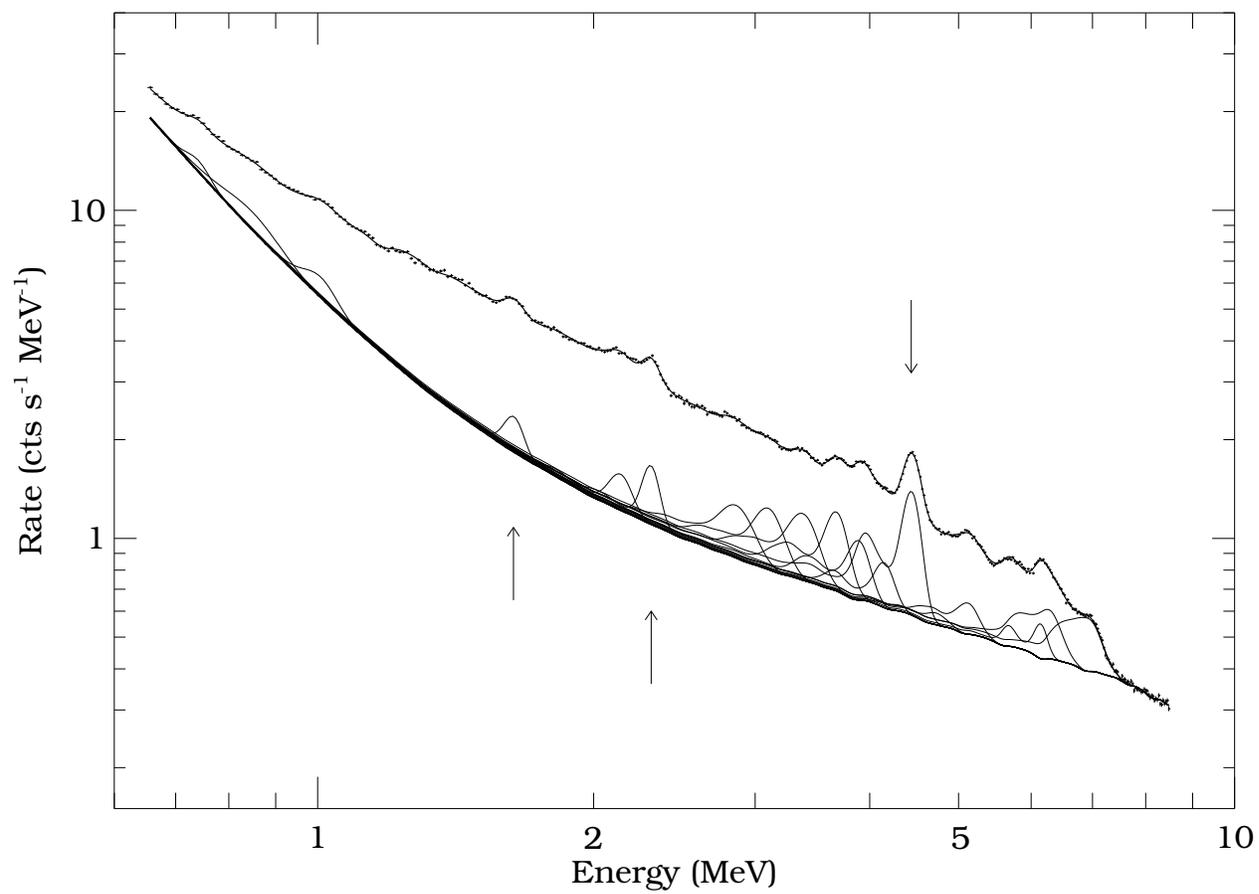
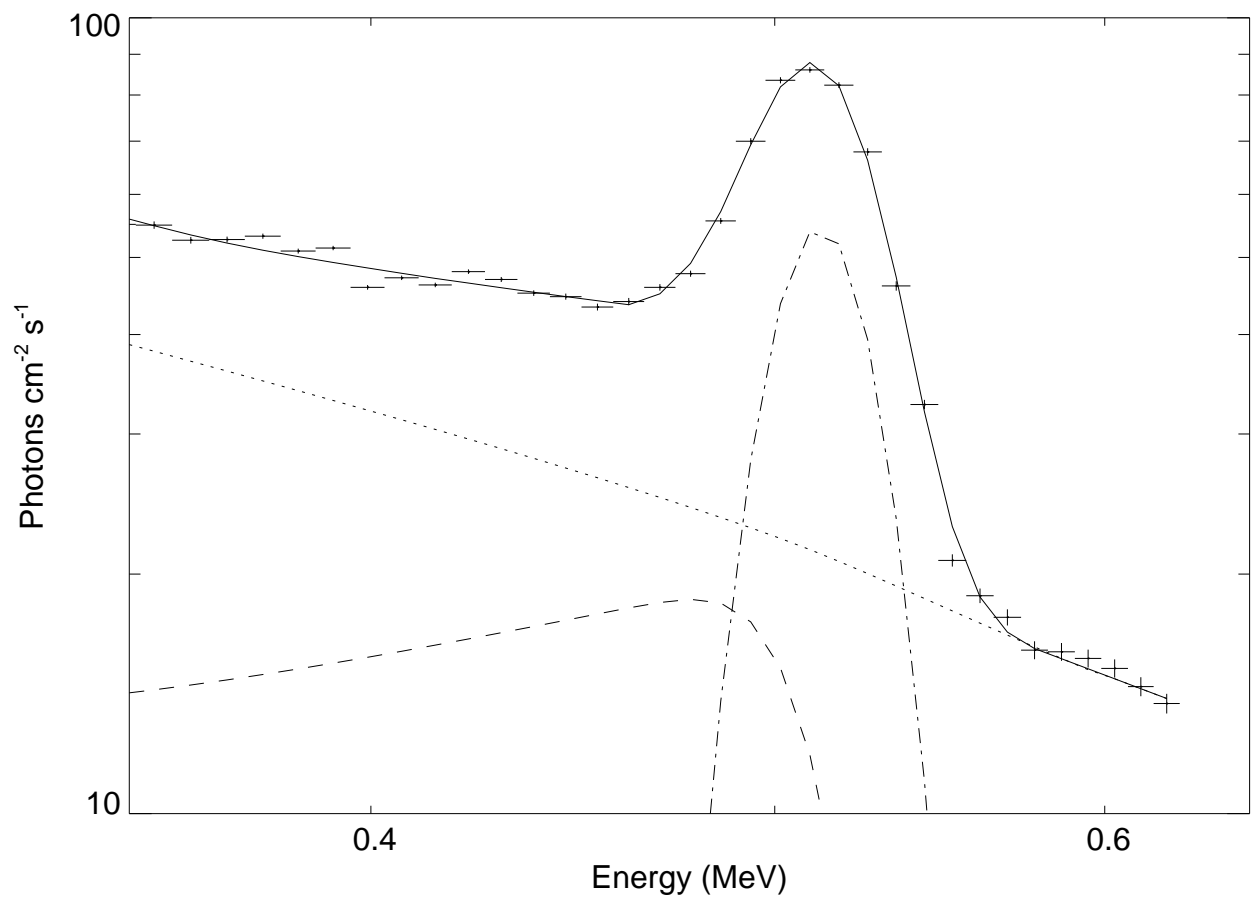
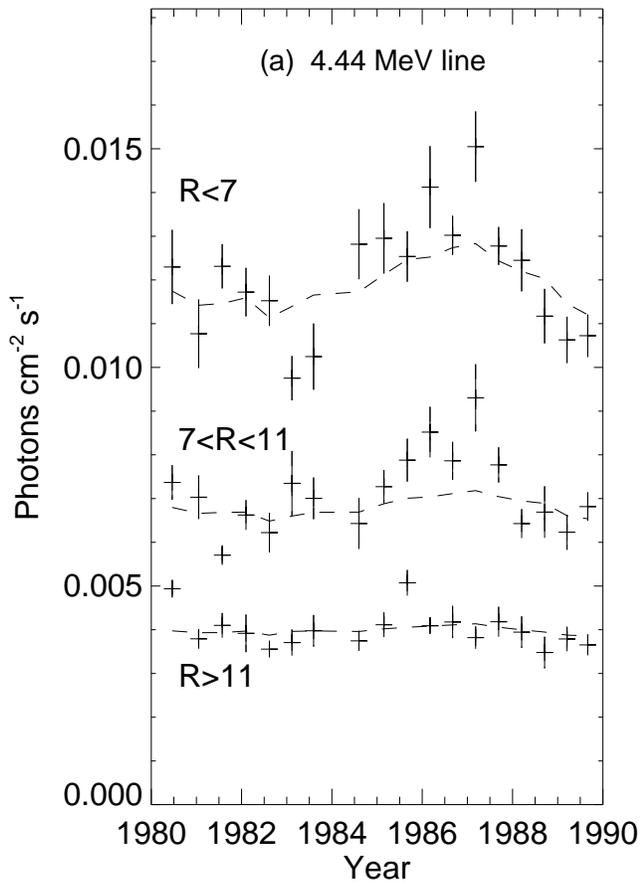
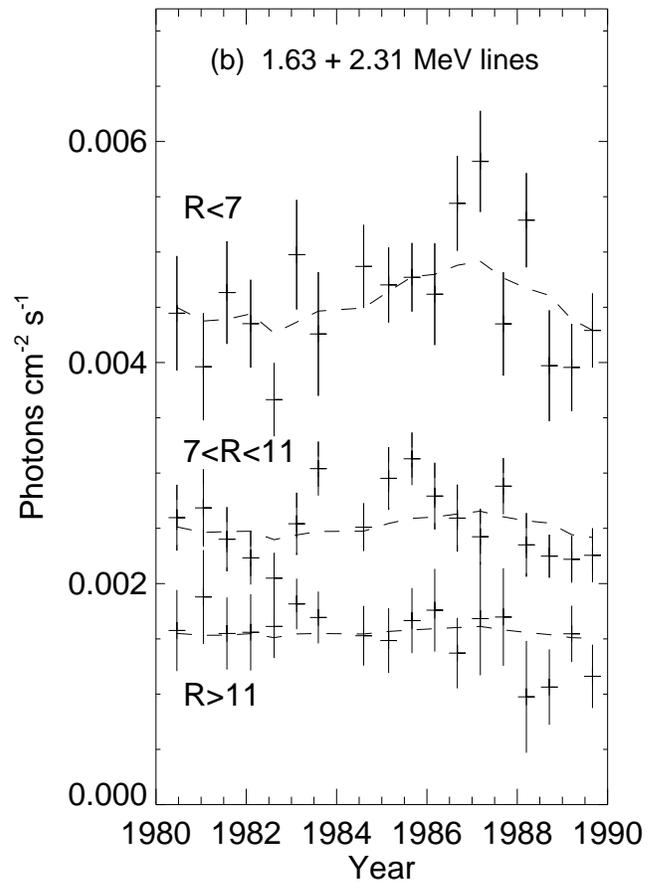
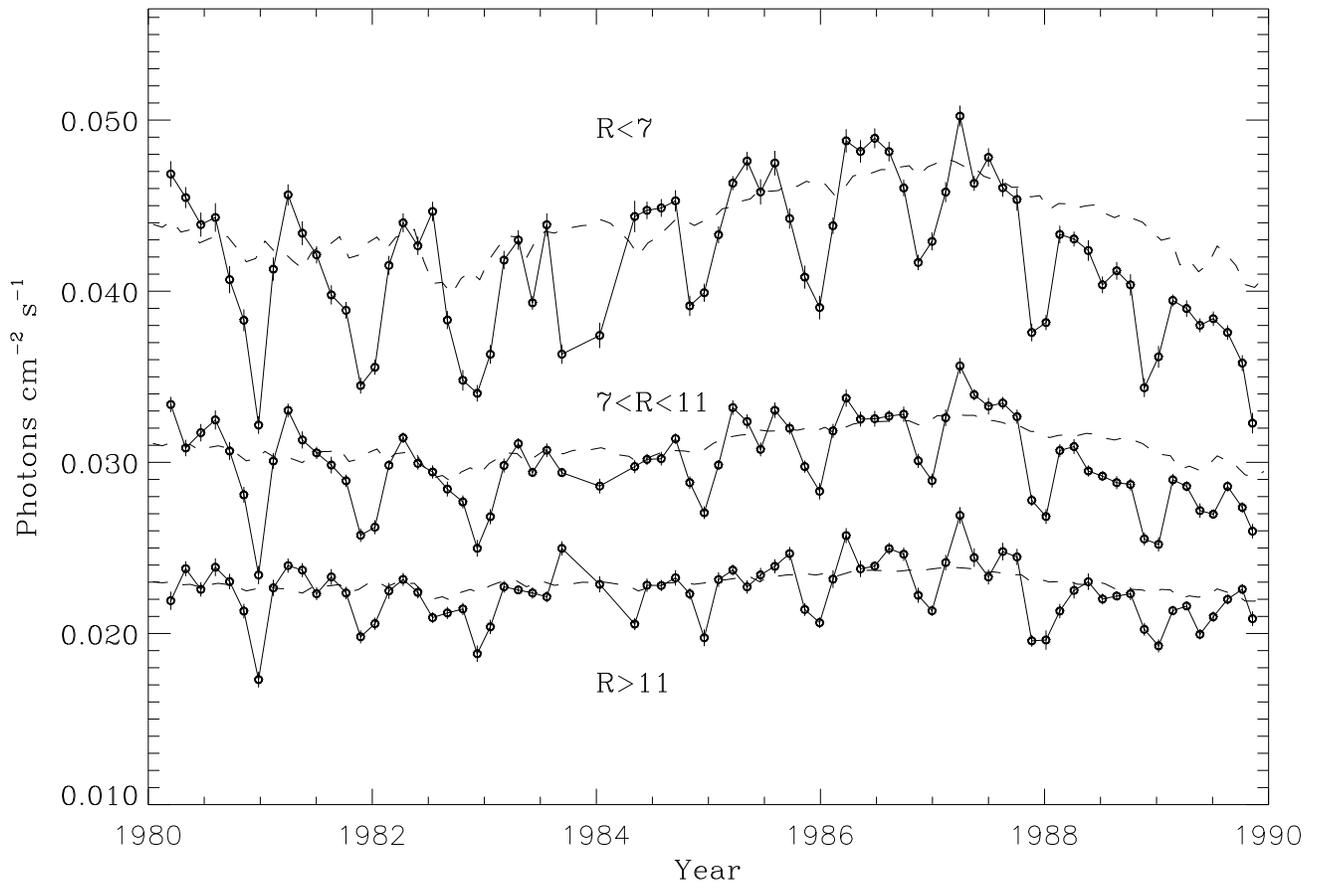
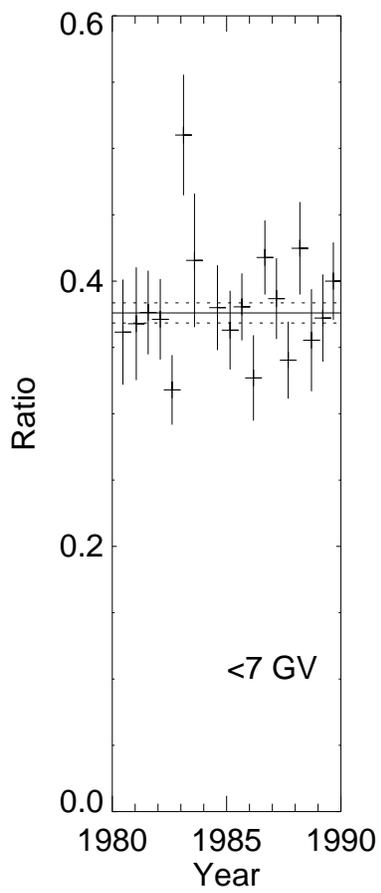
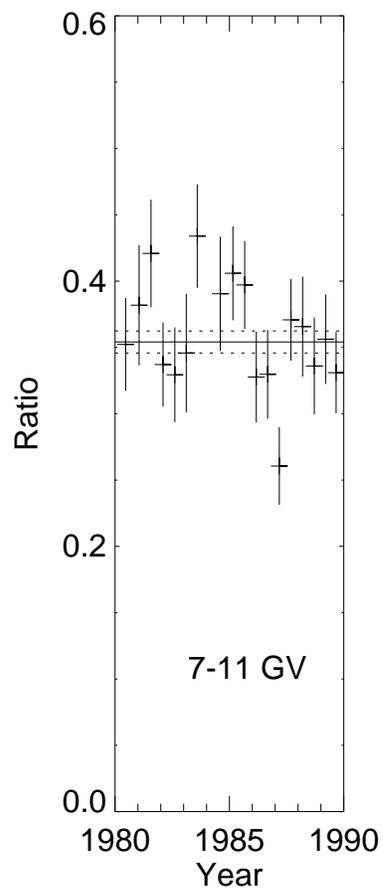
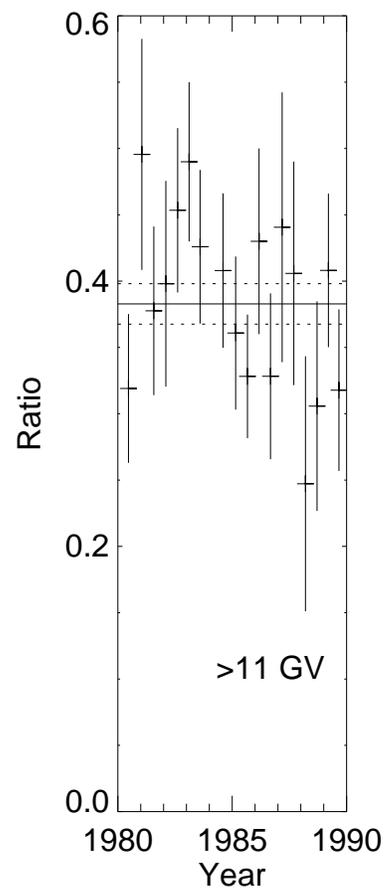
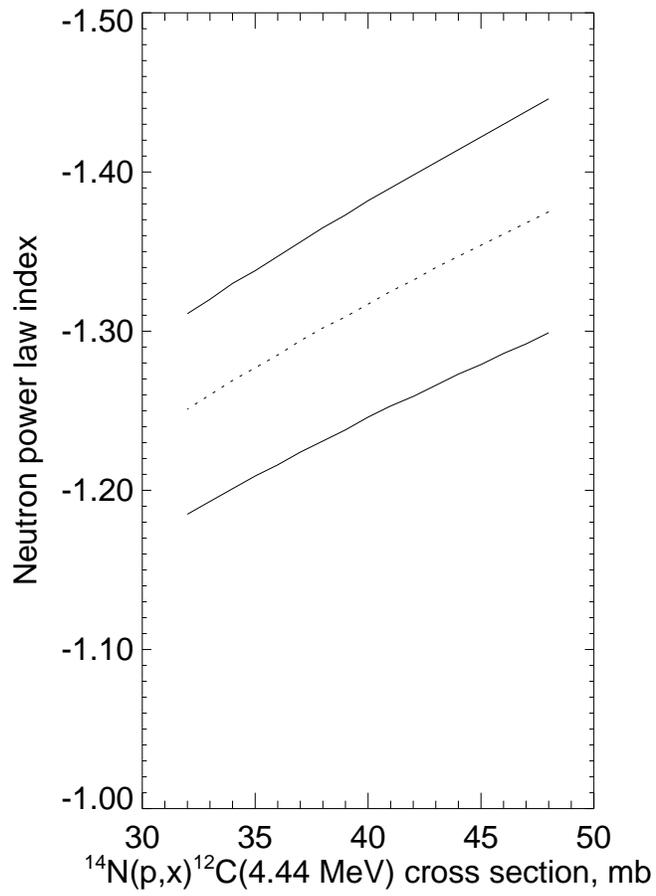
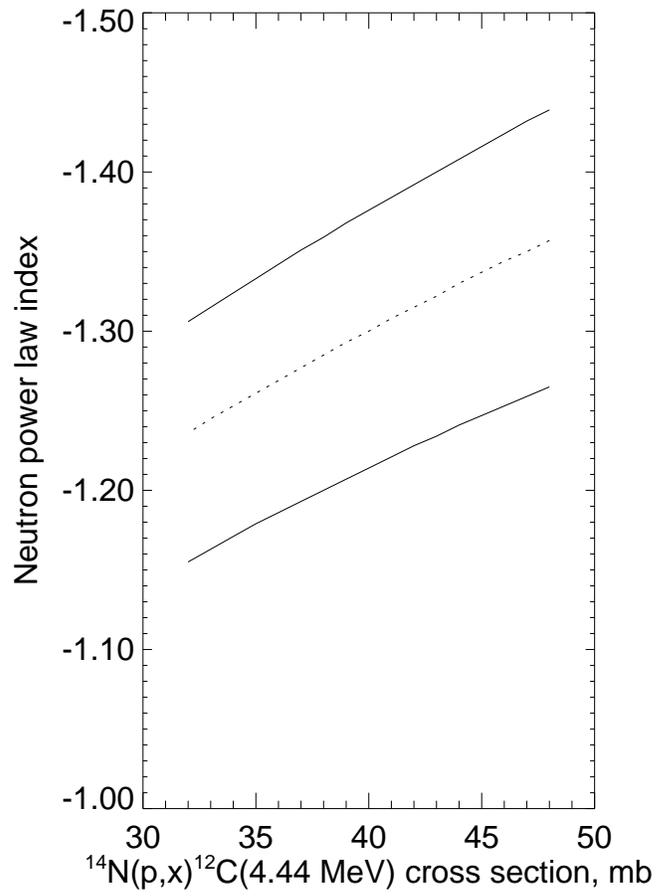
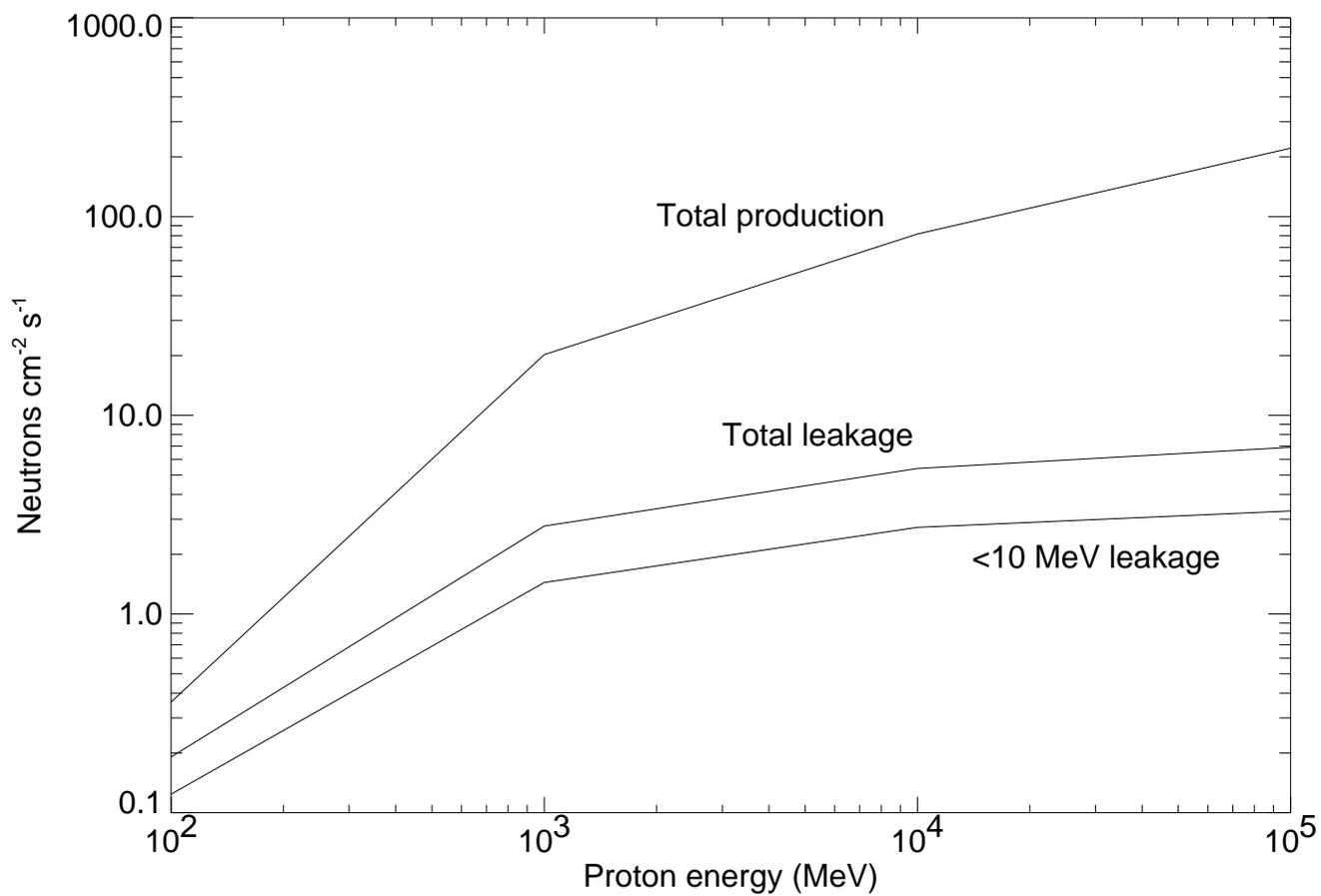
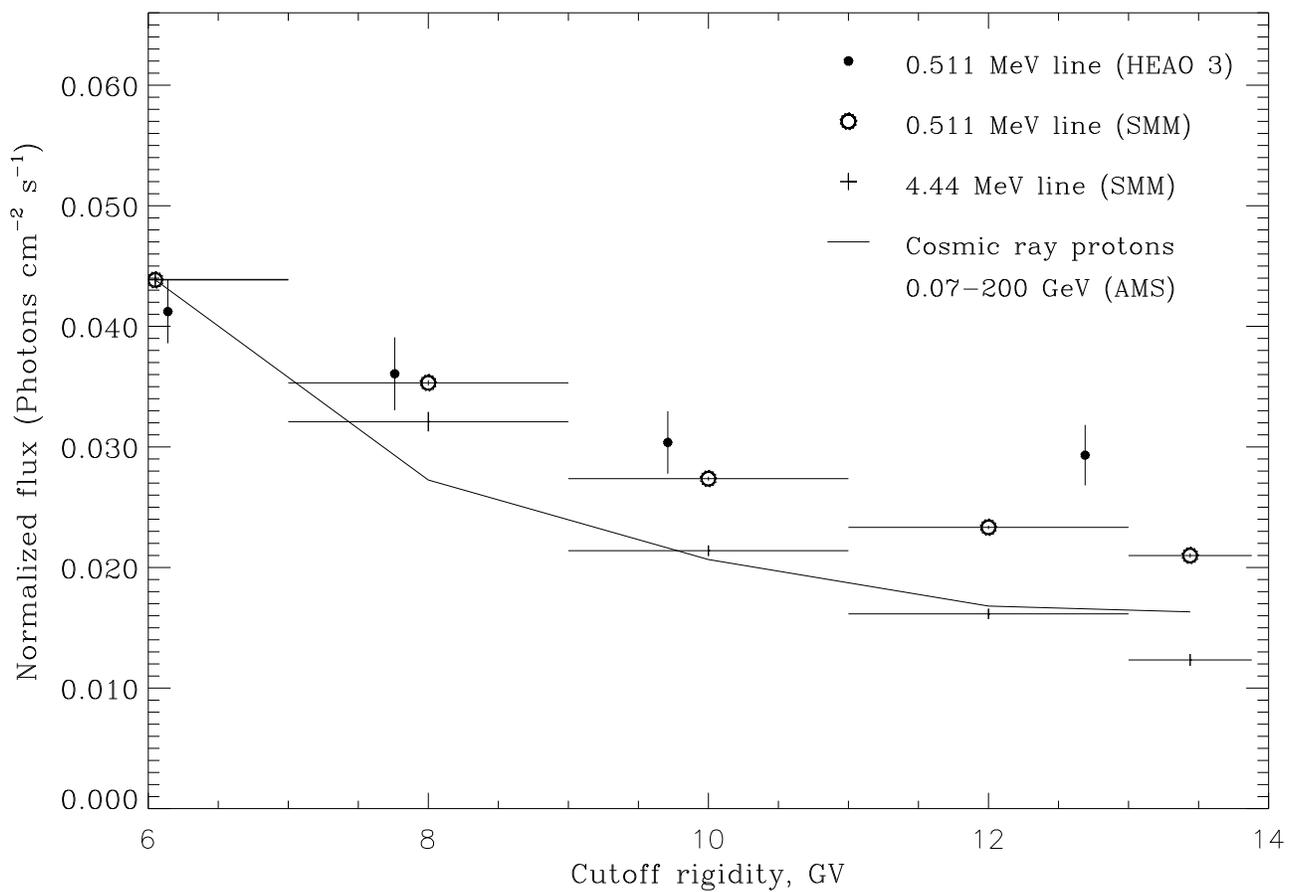
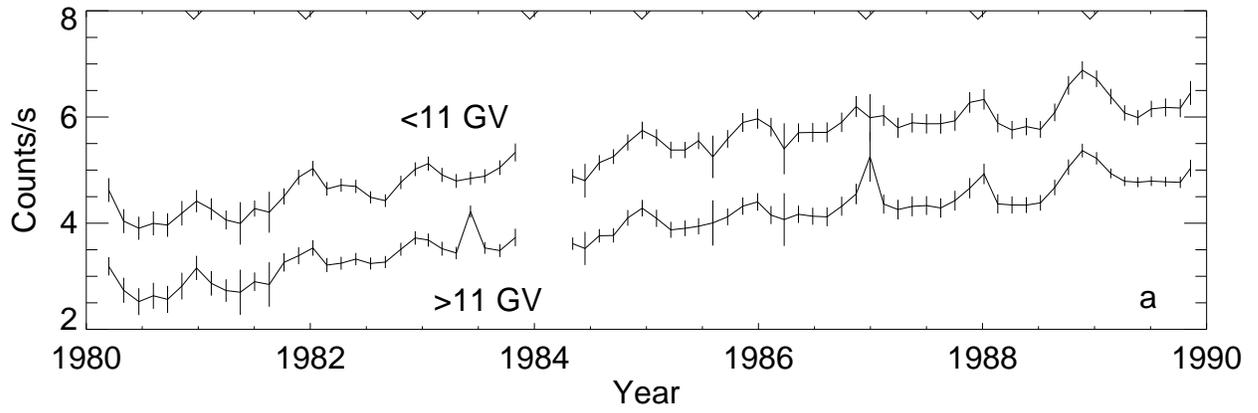
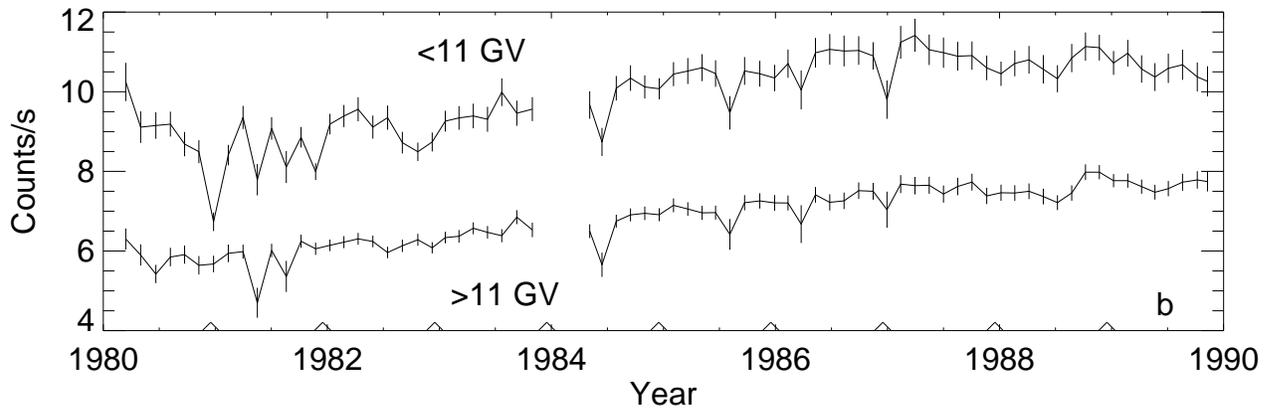
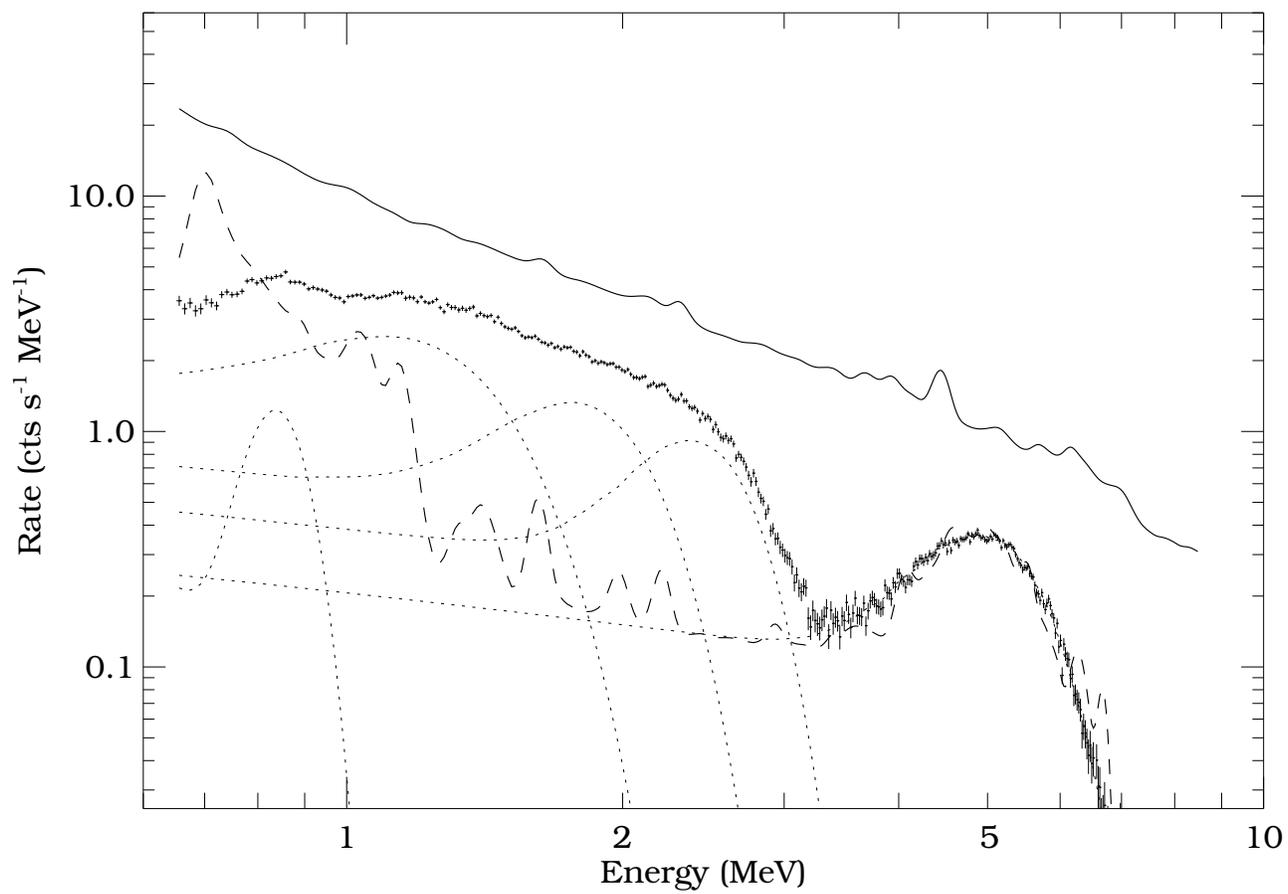
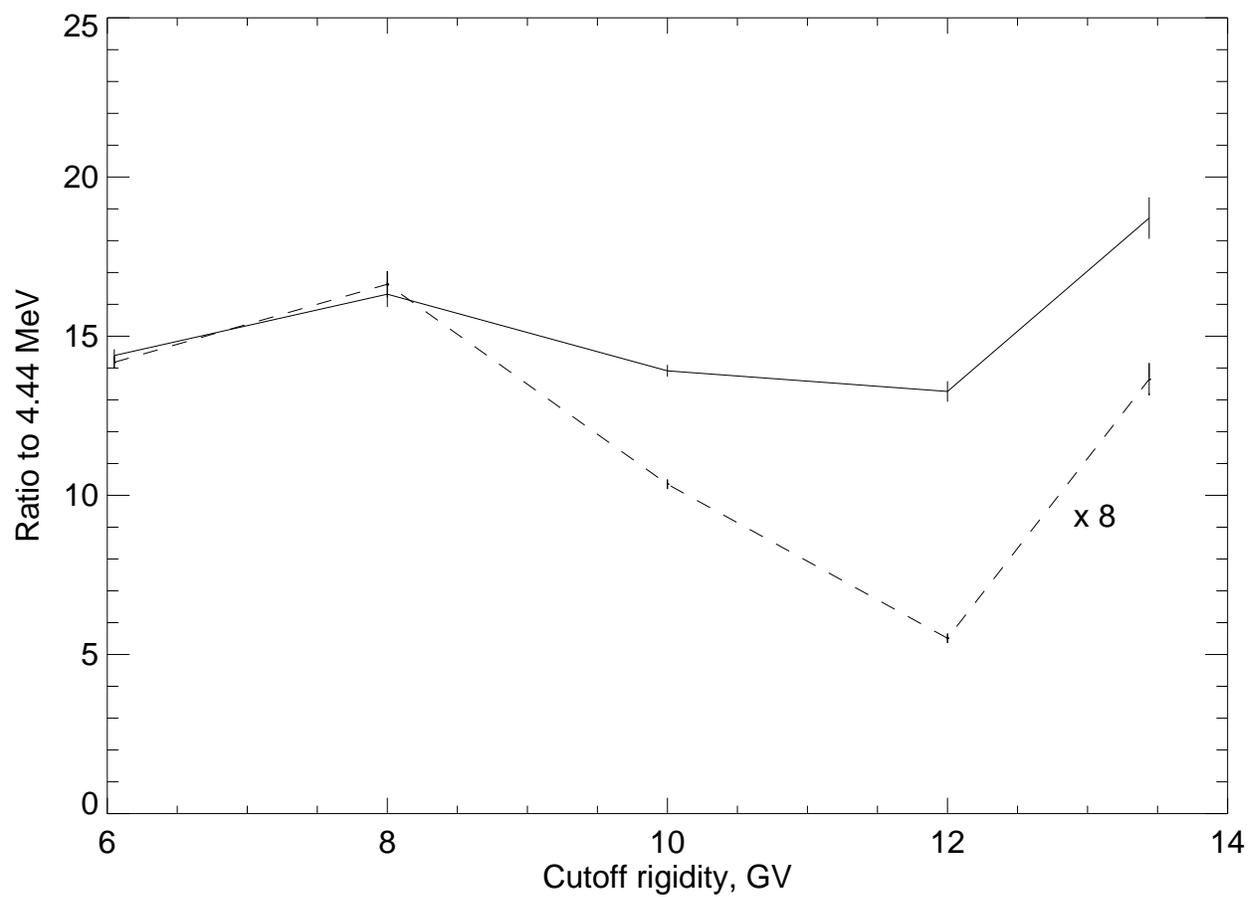
\end{document}